\documentclass[aaspp4]{aastex}
\usepackage{emulateapj5,apjfonts}

\newcommand{\chandra}{{\it Chandra}}
\newcommand{\hst}{{\it HST}}

\newcommand{\eps}{erg s$^{-1}$}
\newcommand{\pcm}{cm$^{-2}$}

\slugcomment{To appear in {\it The Astrophysical Journal.}}

\lefthead{Terashima et al.}
\righthead{Ultraluminous X-ray Sources in M51}

\begin{document}

\title{Hubble Space Telescope Identification of the Optical
Counterparts of Ultraluminous X-ray Sources in M51}

\author{Yuichi Terashima\altaffilmark{1},
Hirohiko Inoue\altaffilmark{1, 2},
 and
Andrew S. Wilson\altaffilmark{3, 4}
}

\altaffiltext{1}{Institute of Space and Astronautical Science, 3-1-1 Yoshinodai, Sagamihara, Kanagawa 229-8510, Japan}

\altaffiltext{2}{Department of Physics, Tokyo Institute of Technology, 2-12-1, Ohokayama, Meguro, Tokyo 152-8551, Japan}

\altaffiltext{3}{Astronomy Department, University of Maryland, College Park, 
MD 20742}

\altaffiltext{4}{Adjunct Astronomer, Space Telescope Science Institute, 3700
San Martin Drive, Baltimore, MD 21218}

\begin{abstract}
We present the results of a search for optical identifications of
ultraluminous X-ray sources (ULXs) in M51 by using mosaic images taken
with the {\it Hubble Space Telescope} Advanced Camera for Surveys in
filters F435W (B), F555W (V), F814W (I), and F658N (H$\alpha$). Our
sample consisting of nine ULXs is defined by analyzing the three {\it
Chandra} observations of M51 performed in 2000 June, 2001 June, and
2003 Aug.  We found four ULXs have one or two candidates for
counterparts, while two have multiple stars within their error
circles. The other three have no candidate counterparts. Four ULXs are
located near or in a star cluster, while others have no association
with a cluster. These results indicate that the companion star,
environment, and origin of ULXs are probably heterogeneous.

\end{abstract}


\section{Introduction}

  A fraction of galaxies contain X-ray sources whose luminosities
exceed $10^{39}$ \eps; these sources are known as ultraluminous
compact X-ray sources (ULXs; Makishima et al. 2000). Their nature is
still under debate, with models including a black hole with a mass
greater than a few times 10$M_{\odot}$ (intermediate mass black hole;
IMBH), super Eddington accretion or beamed emission from a neutron
star or stellar mass black hole (e.g., Miller \& Colbert 2004;
Fabbiano \& White 2005). Optical counterparts of ULXs have the
potential to provide us with significant insights into the nature of
ULXs. If a companion star is detected, it could be used to measure the
mass function of a ULX binary system. The environments of ULXs are
also quite useful for elucidating the origin and formation mechanism
of ULXs.

  Optical counterparts of ULXs have been searched for from the ground
and with the {\it Hubble Space Telescope} (\hst ). One or a few O--B
type stars have been proposed as candidates for the companion star of
some ULXs (e.g., M81, Liu et al. 2002; NGC 5204, Liu et al. 2004; M101
Kuntz et al. 2005), while some ULXs have no obvious stellar
counterparts (e.g., NGC 4559 X10, Cropper et al. 2004). Many ULXs are
located near a star cluster (e.g., Kaaret et al. 2004), or the center
of an optical emission line nebula, which may be a supernova remnant
or ionized by hot stars (e.g., Pakull \& Mirioni 2002; Liu et
al. 2005; see Liu \& Mirabel 2005 for a catalog).  These environments
are suggestive of an episode of recent star formation and a close
connection between the formation of ULXs and stars. Some ULXs reside
in a high-excitation optical emission-line nebula that may be
photoionized by unbeamed X-ray emission from one or more of the ULXs
(Pakull \& Mirioni 2002). Thus the optical counterparts and
environments of ULXs are not uniform and might suggest a heterogeneous
nature for ULXs. Therefore, more systematic searches for ULX
counterparts are of great importance.

%  background AGN

%  star cluster / multiple stars:  Roberts N4559; Liu et al. HST 2005 ...  M82?

  M51 (NGC 5194 and its companion NGC 5195) is an ideal object in
which to search for optical counterparts and to investigate their
environments in a systematic way because of its large number of ULXs
(seven in NGC 5194 and two in NGC 5195; Terashima \& Wilson 2004) and
relative proximity (8.4 Mpc; Feldmeier et al. 1997). The X-ray
luminosities of the ULXs are modest and in a narrow range ($1-4\times
10^{39}$ erg s$^{-1}$). These facts make the M51 ULXs valuable for
studying ULXs with luminosities between ordinary black holes
($<10^{39}$ \eps) and their highest known values ($\sim10^{41}$ \eps ;
M82, Matsumoto et al. 2001; NGC 2276 Davis \& Mushotzky 2004). Liu et
al. (2005) imaged two ULXs (CXOM51~J133001.0+471344 and
CXOM51~J133007.6+471106) with {\hst} WFPC2 in filters F450W, F555W,
and F814W. The former ULX is located on the rim of a star cluster and
there exist a few faint stars in the error circle, while there are
seven stellar objects within the error circle of the latter. There are
no bright stars younger than 10 million years in the error
circles. Liu et al. (2005) estimated that these two ULXs are in
regions with stars younger than $10^{7.8}$ years.

  In this paper, we present candidates for optical counterparts
and environments of all of the nine ULXs in M51 by utilizing the fine
spatial resolution and wide field of view of a recent observation with
the {\hst} Advanced Camera for Surveys (ACS).

\section{Observations and Analysis}

\subsection{Chandra Observations}

% Chandra obs. and ULXs

  A sample of ULXs in M51 was constructed by analyzing three
{\chandra} data sets obtained on 2000 June 20, 2001 June 23--24, and
2003 August 7--8, with effective exposure times of 14.9, 26.8, and
42.9 ksec, respectively. Results from the first and second
observations are presented in Terashima \& Wilson (2004), while we
have newly analyzed the last data obtained by us. Using the first two
data sets, Terashima \& Wilson (2004) found nine ULXs, defined as
off-nuclear X-ray sources each with a luminosity greater than
$10^{39}$ {\eps} in the 0.5--8 keV band. These nine ULXs are shown in
Table 1.  The names for the ULXs are also defined in Table 1 in order
of increasing right ascension. Liu \& Mirabel (2005) assigned the same
numbers from ULX-1 to ULX-7. Liu et al. (2005) used different
definitions: their ULX-3 and ULX-5 are our ULX-7 and ULX-9,
respectively.

% Chandra astrometry

  The three {\chandra} observations were combined to perform an X-ray
source search with the best available photon statistics. Source
detection was performed by using ``wavdetect'' in the CIAO version
3.0.2 software package. Details of the procedures are the same as in
Terashima \& Wilson (2004). The resulting source list was compared
with source positions obtained in other wavebands. One optical/near
infrared star and the radio nucleus of NGC 5194 were used to estimate
the accuracy of astrometry and we found their source positions agree
to within $\approx$ 0.2 arcsec with the Chandra X-ray positions (the
near infrared position of the star was taken from the 2MASS catalog,
while the radio position of the nucleus was taken from a VLA
observation in A-configuration (Hagiwara et al. 2001). Since the
directions of the offsets of source positions were not systematic, we
made no corrections to the coordinates. The newly-determined positions
of the ULXs are shown in Table 1. The new positions are in good
agreement with the previous measurement, and differences are typically
0.1 arcsec or less. The statistical errors of the positions of the
ULXs are 0.07 arcsec or better.

\subsection{{\em HST} Observations}

% HST observation

  The M51 system was mapped with the {\hst} ACS in 2005 Jan. as a part
of the Hubble Heritage program. The data consist of six ($2\times3$)
images in the four filters F435W (B), F555W (V), F814W (I), and F658N
(H$\alpha$), with exposure times of 2720, 1360, 1360, and 2720 sec,
respectively. We use the drizzle-combined fits images (the version 1.0
mosaics) taken from the archives in the following analysis. We
measured the positions and fluxes of detected sources by point-spread
function fitting to the images using the {\tt daophot} package.

%  [In searching for optical counterparts of ULXs, accuracy of relative
%astrometry between the HST and Chandra images are essential.]

  In order to improve the astrometry of the {\hst}, we compared
point-like sources in the ACS images with sources listed in the 2MASS
catalogue and found systematic offsets of 0.1 arcsec and 0.7 arcsec in
the right ascension and declination directions, respectively. We
shifted the ACS images to the 2MASS positions. After this alignment,
we searched for coincident point-like sources detected in both the ACS
and {\chandra} images. Four X-ray sources (CXOM51~J132938.6+471336,
CXOM51~J132938.9+471324, CXOM51~J132943.4+471525, and
CXOM51~J133011.0+471041) have point-like counterparts in the ACS
images. All four sources are detected in the three bands B, V, and I.
These sources are relatively bright in X-rays and have 130--310 counts
in the 0.5--8 keV band in the combined data of the three observations.
The statistical errors of their {\chandra} positions are 0.05--0.12
arcsec. The rms of the position differences between {\chandra} and
{\hst} are 0.09 and 0.10 arcsec in the directions of right ascension
and declination, respectively. Thus, we adopt 0.1 arcsec as the
uncertainty between the source locations in the optical and X-ray
images. We use a one sigma uncertainty of 0.14 arcsec by combining
both this uncertainty (0.1 arcsec) and the X-ray position error of the
ULXs (better than 0.1 arcsec) in quadrature. If we assume the
probability distribution of a source location is a 2-dimensional
Gaussian, this uncertainty corresponds to a 90\% confidence error
radius of 0.3 arcsec (12 pc), which is used throughout this paper.

\section{Results}

\subsection{X-ray Spectra and Variability}

  Among the nine ULXs found in the observations in 2000 and 2001, only
four (ULX-2, 3, 7, and 9) fulfil the definition of a ULX in the data
obtained in 2003. The luminosities of three other ULXs (ULX-1, 5, and
8) decreased to $(7-9)\times 10^{38}$ {\eps} and two objects (ULX-4
and ULX-6) diminished to below 1.0$\times10^{37}$ and
$9.2\times10^{36}$ {\eps} in the 0.5--8 keV band (95\% confidence
upper limit), respectively, where we assumed a power law spectrum with
a photon index of 2.0 modified by the Galactic absorption ($N_{\rm
H}=1.4\times10^{20}$ cm$^{-2}$). No new ULXs appeared in the 2003
observation. It is interesting to note that the two objects ULX-4 and
ULX-6 were detected only in the 2001 observation and that their
spectra are completely different: a power law with a photon index of
1.55 and an MCD with an inner disk temperature of 0.10 keV.

  We modelled the X-ray spectra of the seven sources detected in 2003
using the XSPEC spectral fitting package. The spectra were grouped so
that each spectral bin contained at least 20 counts to allow use of
the $\chi^2$ minimization technique. Background spectra were taken
from a source-free region near the source and subtracted from the
on-source spectra. Two continuum models, a power law and a multicolor
disk blackbody (MCD; Mitsuda et al.  1984) modified by photoelectric
absorption, were examined.  The results of spectral fitting for the
third {\it Chandra} observation (in 2003) are summarized in Table 2,
while the results for the 2000 and 2001 observations are presented in
Terashima \& Wilson (2004). The observed fluxes and absorption
corrected luminosities are also shown. The spectra of three ULXs
(ULX-5, 7, and 9) are better fitted with a power law than an MCD
model. ULX-2 shows a very soft spectrum and an MCD fits the data
better. We applied a blackbody model to the spectrum of ULX-2 and
obtained a similar quality of fit. The result is also shown in Table
2. Other ULXs are well fitted with either a power law or an MCD model.

  We examined the light curves of the seven ULXs detected in all three
observations. ULX-1, 2, 7, and 8 showed mild variability. Their
intensities changed by a factor of 1.5--2 in several thousand to ten
thousand seconds.  A two hour periodicity in ULX-7 reported in Liu et
al. (2002) was not confirmed. Only two (or maybe three) periods are
seen at the beginning of the observation, but the periodicity is not
clear in the latter half.

\subsection{Optical Counterparts}

% presence/absence of candidate stellar counterparts

  Fig. 1 shows a true-color image around the nine ULXs. There are one
or two candidate stellar counterparts in the error circles of ULX-1,
2, 8, and 9. The apparent magnitudes of these stars are shown in Table
3. A very faint star may be present as well within the error circles
of ULX-1 and ULX-2. There are no dense or large star clusters near
ULX-1, but ULX-1 may be in a shell with an oval shape seen in an
H$\alpha$ image (Fig. 2). The diameter of the structure is about 260
pc. The H$\alpha$ luminosity of this structure is $4.9\times10^{36}$
erg s$^{-1}$, which includes red-shifted [\ion{N}{2}]$\lambda$6583
located in the filter bandpass.  ULX-2 is located near the rim of a
large star cluster. The projected distance between ULX-2 and the
center of this cluster is about 170 pc. ULX-8 is in the tidal tail
connecting NGC 5194 and NGC 5195 and located 2.8 kpc east of the
nucleus of NGC 5195. There are two very faint sources in the error
circle in the B band image. No H$\alpha$ emitting region is seen
around this ULX. There is one source in the error circle of ULX-9.
This counterpart corresponds to ``c1'' presented in Liu et
al. (2005). ULX-9 seems not to be associated with a cluster: the
projected distance to the nearest cluster is 240 pc. It is worth
noting that this distance is larger than a trend suggested by
observations that X-ray binaries with a higher luminosity tend to be
located near a star cluster (Kaaret et al. 2004). Images of the
$2.5''\times2.5''$ region around the ULXs with possible counterparts
are shown in Fig. 3.

% [Photometry of individual candidates]

  We measured the apparent magnitude and colors of the candidate IDs
in the STMAG system. Results are summarized in Table 3.  The amount of
the Galactic extinction derived from \ion{H}{1} maps is negligible
($A_{\rm V}\approx$ 0; Burstein \& Heiles 1984) and no correction was
made. These magnitudes are converted to the Johnson system by
referring to Sirianni et al. (2005), and color-magnitude diagrams are
plotted in Fig. 4 along with the evolutionary tracks with $Z=0.02$ of
Lejeune \& Schaerer (2001).  $Z=0.02$ corresponds to the solar
metallicity [Fe/H] = 0. The magnitude and colors of the candidate
counterparts of ULX-1, ULX-2, and ULX-9 are consistent with B2-8,
O5-B2, and F2-5 supergiants, respectively. The B-band magnitudes of
the two candidate counterparts to ULX-8 are consistent with B2V type.

  %%% Comparison with QSO colors

  The colors of the candidate IDs were compared with those of quasars
to test whether the optical sources are consistent with background
AGNs. We used 38349 quasars at a redshift $z<2.1$ in the Sloan digital
sky survey third data release (Schneider et al. 2005). Photometric
data taken at the $g,$ $r,$ and $i$ bands were converted to the colors
$B-V$ and $V-I$ by using the conversion factors appropriate for
$z<2.1$ given in Table 1 of Jester et al. (2005). None of the quasars
have colors consistent with the candidate counterparts of ULX-1 and
ULX-2, while the color of the ULX-9 counterpart is in the range of the
quasars. The expected number of background AGNs with a flux in the
0.5--8 keV band greater than $1.2\times10^{-13}$ {\eps}{\pcm} (i.e.,
$1.0\times10^{39}$ {\eps} at 8.4 Mpc) coincidently detected within the
areas of NGC 5194 and NGC 5195 is 0.24 and 0.021, respectively, if the
$\log N$-$\log S$ relation in the 2--10 keV band shown in Fig. 3 of
Ueda et al. (2003) and a photon index of 1.8 are assumed.

 We estimated the probability that these counterparts are chance
coincidences with an unrelated star. An error circle with a radius of
0.3 arcsec was randomly placed in the region around the ULXs and we
calculated the probability of detection of one or more stars within
the circle. Since the probabilities are a function of the magnitude of
the stars, the following two cases were examined: detection of stars
of any brightness detected in the ACS observation and of brightness
greater than or equal to that of the proposed ID. The results are
shown in Table 3.

  The two ULXs ULX-4 and ULX-7 are in a region with a relatively high
stellar density in or near a star cluster. ULX-4 is located near the
center of a star cluster, and there are many stars within the error
circle. ULX-7 is near the rim of a cluster and there are at least four
stars in the error circle.

  There are no obvious optical counterparts for ULX-3, 5, and 6.
ULX-3 is located at the position of one of the dark lanes in the inner
spiral arms. Its large X-ray absorbing column density, $N_{\rm H}
\approx 5\times 10^{22}$ cm$^{-2}$ (Terashima \& Wilson 2004), is
qualitatively consistent with the location. Thus, extinction to this
object is likely to be large, and our data cannot rule out the
presence of optically luminous counterparts. A hint of the presence of
a very faint star may be a counterpart suffering from heavy
extinction.  This region is bright in H$\alpha$. ULX-5 is located
outside the outer arm. The stellar density around there is relatively
small and there are no bright stars. An extended source visible in all
the optical bands is present 80 pc west of ULX-5. An H$\alpha$ image
of the extended source consists of a ring-shaped structure with a
radius of about 50 pc and a bright spot at around (13:29:53.55,
+47:14:35.6) (J2000.0). It is not clear whether this object is related
to the ULX or not. ULX-6 is in the bright stellar envelope of NGC
5195. Individual point sources are not clearly seen in this region.
In summary, there is no clear association with a star or star cluster
for ULX-3, 5 or 6.

%[presence/absence of Halpha nebulosity]

  Association with a star cluster is of significant importance for
understanding the environment of ULXs. Among the nine ULXs, four
(ULX-2, 4, 7, and 9) are located near the center or rim of a star
cluster. Their H$\alpha$ images show diffuse emission surrounding the
clusters. These ULXs are located around the rim of the extended
emission.

\section{Discussion}

  The ACS images show that the ULXs in M51 can be classified into
three subclasses in terms of their optical counterparts: (1) there are
one or two stars in the error circle, (2) the ULX is located in a
region with relatively high stellar density and multiple candidate
counterparts are present, and (3) there are no candidate counterparts.

% companion star
% presence/absence of a OB star.

  The spectral types of the candidate companion stars of ULX-1 and
ULX-2 are as early as OB-type and indicate that these systems are
high-mass X-ray binaries. Candidate companions of such early type
stars are also reported for ULXs in other galaxies. The counterpart of
ULX-9 has a later spectral type (F-type). A comparison with the
isochrones for $Z$=0.02 (Lejeune \& Schaerer 2001) shows that the
counterparts of ULX-1 and ULX-2 have an age of $10^{6.6-7.2}$ yrs,
while that of the counterpart of ULX-9 is around $10^{7.6-7.8}$ yrs.
Some other ULXs also have lower mass companions; a main sequence star
with the limiting magnitude of these observation ($M_{\rm B}=0.4$) or
fainter corresponds to A- or later type. Thus, there exist both
high-mass and low-mass companions among the ULXs in M51.

% m_B = 27.1, 28.3, and 27.1
% M_B = -2.5, -1.3, -2.5
% distance modulus = 29.621

% limiting mag 27.3

%  magnitude --- single star or unresolved multiple stars?

% presence/absence of star cluster

  The ULXs in our sample are in two environments: (1) ULXs located in
or near a star cluster or associated with an \ion{H}{2} region, and
(2) no connection to a star cluster. This fact suggests that the
presence of a star cluster is not always required to form a ULX. If
ULXs not associated with a star cluster do contain an IMBH, IMBH could
be formed by a process independent of recent star formation or a star
cluster currently observed.  In this case, an IMBH may be a remnant of
past active star formation or a population III star (Madau \& Rees
2001). Alternatively, the compact object in ULXs could have a stellar
mass expected from the known evolution of ordinary stars. If this is
the case, the observed X-ray luminosities require only mild (a factor
of a few) super-Eddington luminosities for a 10$M_{\odot}$ black hole,
for example. Such a luminosity is expected by some workers from an
accretion disk with a high accretion rate (e.g., Ohsuga et
al. 2003). Thus the wide variety of the environments may suggest a
heterogeneous ULX formation history and/or black hole mass at least
for ULXs with a moderate luminosity (a few $\times10^{39}$ {\eps}).

\acknowledgements

This research was partially supported by NASA through LTSA grant NAG 513065
to the University of Maryland.

\clearpage

\begin{table}[t]
\tabletypesize{\small}
\begin{center}
\caption{Ultraluminous Compact X-ray Sources in M51}
\begin{tabular}{ccccc}
\hline \hline
ULX Number & CXO Name$^a$   & \multicolumn{2}{c}{X-ray Position$^b$} & \\
           & CXOM51     & R.A. (J2000.0)  & Dec. (J2000.0)   & \\ 
\hline
ULX-1 & J132939.5+471244 & 13 29 39.47 & +47 12 43.60 \\
ULX-2 & J132943.3+471135 & 13 29 43.31 & +47 11 34.80 \\
ULX-3 & J132950.7+471155 & 13 29 50.68 & +47 11 55.21 \\
ULX-4 & J132953.3+471042 & 13 29 53.31 & +47 10 42.46 \\
ULX-5 & J132953.7+471436 & 13 29 53.72 & +47 14 35.74 \\
ULX-6 & J132958.4+471547 & 13 29 58.38 & +47 15 47.40 \\
ULX-7 & J133001.0+471344 & 13 30 01.01 & +47 13 43.93 \\
ULX-8 & J133006.0+471542 & 13 30 06.00 & +47 15 42.30 \\
ULX-9 & J133007.6+471106 & 13 30 07.55 & +47 11 06.11 \\
\hline
\end{tabular}
\tablecomments{
$a$: {\it Chandra} name presented in Terashima \& Wilson (2004).
$b$: X-ray positions newly determined in this paper.
}
\end{center}
\end{table}

\vspace{1cm}

\begin{table}[th]
\tabletypesize{\small}
\begin{center}
\caption{X-ray Spectral Parameters}
\begin{tabular}{ccccccc}
\hline \hline
ULX Name & $N_{\rm H}$ & Photon index & $kT_{\rm in}$ & Flux$^a$ & Luminosity$^b$ & $\chi^2$/dof \\
           & ($10^{22}$ cm$^{-2}$) &              & (keV)         & ($10^{-14}$ erg s$^{-1}$cm$^{-2}$) & (10$^{39}$erg s$^{-1}$)\\
\hline
ULX-1 & $0.18^{+0.07}_{-0.08}$ & $1.99^{+0.29}_{-0.22}$ & ... & 6.5 & 0.70 & 16.1/15 \\
      & $0.027^{+0.053}_{-0.027}$ & ... & $1.09^{+0.28}_{-0.22}$ & 5.6 & 0.49 & 17.8/15 \\
ULX-2 & $0.58^{+0.12}_{-0.15}$ & ... & $0.092\pm0.003$           & 3.7 & 9.7 & 29.9/28 \\
      & $0.50^{+0.16}_{-0.17}$ & ... & $0.085^{+0.012}_{-0.020}$ & 3.7 & 6.0 & 25.9/28$^c$\\
ULX-3 & $3.8^{+1.1}_{-1.0}$    & $1.52^{+0.54}_{-0.41}$ & ...    & 19.0& 3.3 & 48.0/37 \\
      & $3.1\pm0.7$            & ... & $2.9^{+2.2}_{-0.9}$       & 18.6& 2.5 & 49.1/37 \\
ULX-4 & 0.014$^d$              & 2.0$^d$                & ..     & $<0.12^e$ & $<0.010^e$ & ...\\
ULX-5 & $0.19^{+0.03}_{-0.06}$ & $1.73^{+0.13}_{-0.17}$ & ...    & 8.5 & 0.79& 25.2/22 \\
      & $0.050$                & ...                    & 1.4    & ... &  ...& 34.3/22 \\
ULX-6 & 0.014$^d$              & 2.0$^d$                & ...    & $<0.11^e$ & $<0.0092^e$  & ...\\
ULX-7 & $0.16^{+0.04}_{-0.03}$ & $1.49^{+0.13}_{-0.12}$ & ...    & 32.7& 3.2 & 42.3/49 \\
      & $0.049^{+0.021}_{-0.018}$ & ... & $1.86^{+0.27}_{-0.23}$ & 30.3& 2.7 & 60.9/49\\
ULX-8 & $0.15^{+0.05}_{-0.06}$ & $1.76^{+0.13}_{-0.19}$          & ... & 7.1 & 0.71& 8.4/18\\
      & $0.022^{+0.042}_{-0.022}$ & ... & $1.34^{+0.36}_{-0.27}$ & 6.2 & 0.53& 11.9/18\\ 
ULX-9 & $0.25\pm0.03$  & $2.26\pm0.13$  & ...                    & 24  & 3.0 & 79.2/54\\
      & $0.075$  & ... & $0.97$                                  & ... & ... & 118.3/54\\
\hline
\end{tabular}
\tablecomments{
$a$: Observed flux in the 0.5--8 keV band.
$b$: Luminosity corrected for absorption in the 0.5-8 keV band.
$c$: Blackbody model fit.
$d$: Assumed parameters.
$e$: 95\% confidence upper limit.
}
\end{center}
\end{table}

%ULX-1   src12
%ULX-2   src17
%ULX-3   src30 (emission line)
%ULX-4   ...
%ULX-5   src38
%ULX-6
%ULX-7   src73
%ULX-8   src88
%ULX-9   src91

\begin{table}[t]
\tabletypesize{\small}
\begin{center}
\caption{Optical positions and magnitude}
\begin{tabular}{ccccccccc}
\hline \hline
Name  & \multicolumn{2}{c}{Optical Position} & \multicolumn{3}{c}{Magnitude$^a$}& Chance probability$^b$\\
      & R.A. (J2000.0)  & Dec. (J2000.0)    & B     & V     & I \\ 
\hline
ULX-1 & 13 29 39.467 & +47 12 43.55 & 22.59 & 23.27 & 24.80 & 0.004 (0.48)\\
ULX-2 & 13 29 43.309 & +47 11 34.73 & 23.20 & 24.01 & 25.50 & 0.17 (0.79) \\
ULX-8 & 13 30 06.001 & +47 15 42.28 & 26.51 & ...    & ...  & 0.29 (0.29) \\
      & 13 30 05.977 & +47 15 42.12 & 26.03 & ...    & ...  & ... \\
ULX-9 & 13 30 07.547 & +47 11 06.07 & 25.17 & 25.49 & 26.04 & 0.27 (0.47) \\
\hline
\end{tabular}
\tablecomments{
($a$) Magnitude in the STMAG system, defined as $-2.5\times \log_{10}F -21.10$,
where $F$ is a measured
flux in units of erg cm$^{-2}$ s$^{-1}$ A$^{-1}$.
($b$) Probability that an unrelated star is detected in the
error circle. Probabilities are for stars of brightness greater than
or equal to that of the proposed ID. Probabilities for stars of any
brightness detected by the ACS observation are shown in parentheses.}
\end{center}
\end{table}

\clearpage

\begin{figure}
\plotone{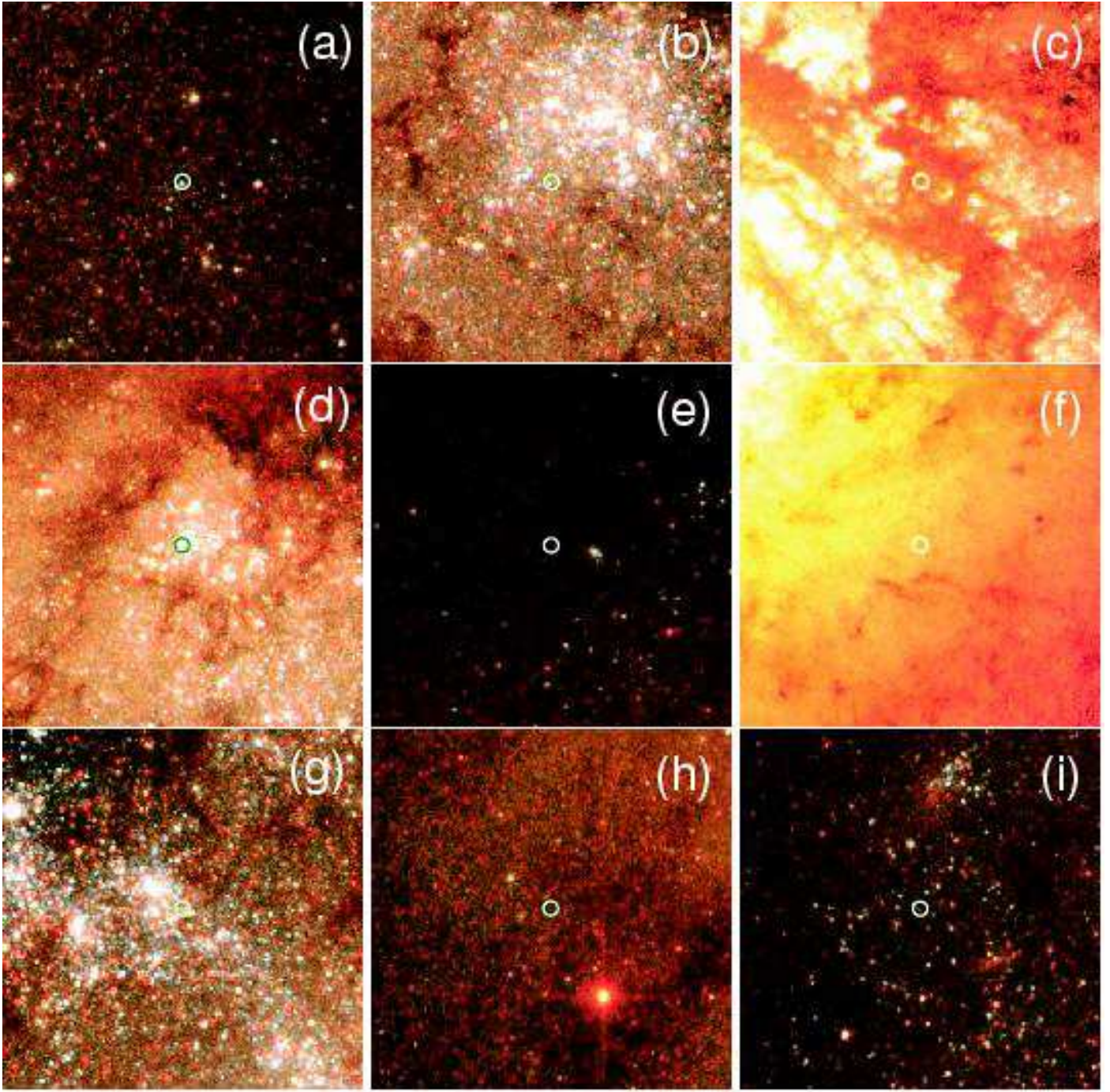}
\caption{ACS true-color images around the nine ULXs in M51. Red, green,
and blue correspond to the F435W, F555W, and F814W filters,
respectively.  The adopted error circles
of ULXs, with a 0.3'' radius,
is shown. The size of each image is $15''\times 15''$ or 610 pc
$\times$ 610 pc. North is up and east is to the left. Fig 1a--1i
correspond to ULX-1 to ULX-9, respectively.}
\end{figure}

\begin{figure}
\begin{center}
\includegraphics[scale=0.37,angle=0]{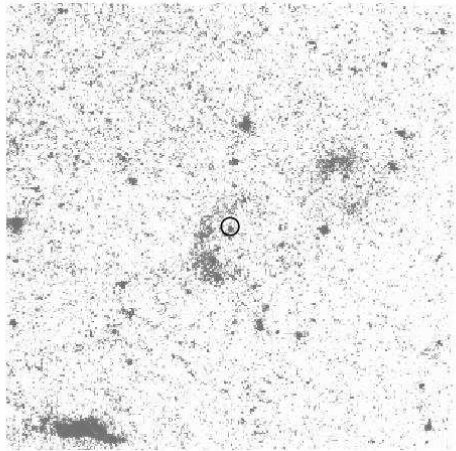}
\figcaption[]{ACS F658N (H$\alpha$) images around ULX-1. Image size is same as Fig. 1. North is up and east is to the left.}
\end{center}
\end{figure}

\begin{figure}
\begin{center}
%\plotone{f3.ps}
%\includegraphics[scale=0.86,angle=0]{f3.ps}
\includegraphics[scale=1.45,angle=0]{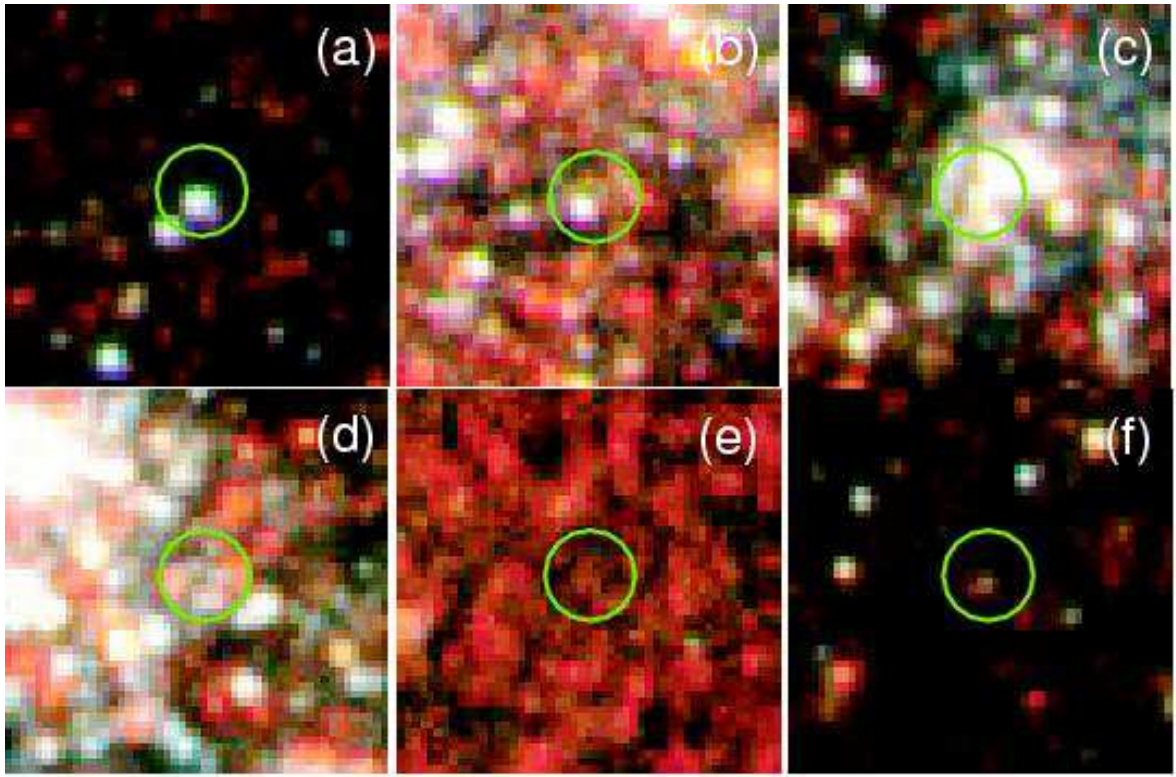}
\figcaption[]{ACS true-color images of the $2.5''\times2.5''$ region around
the ULXs with possible
counterparts. The colors and error circle are 
same as Fig. 1. Fig 3a--3f correspond to ULX-1, 2, 4, 7, 8, and 9,
respectively.}
\end{center}
\end{figure}

\begin{figure}
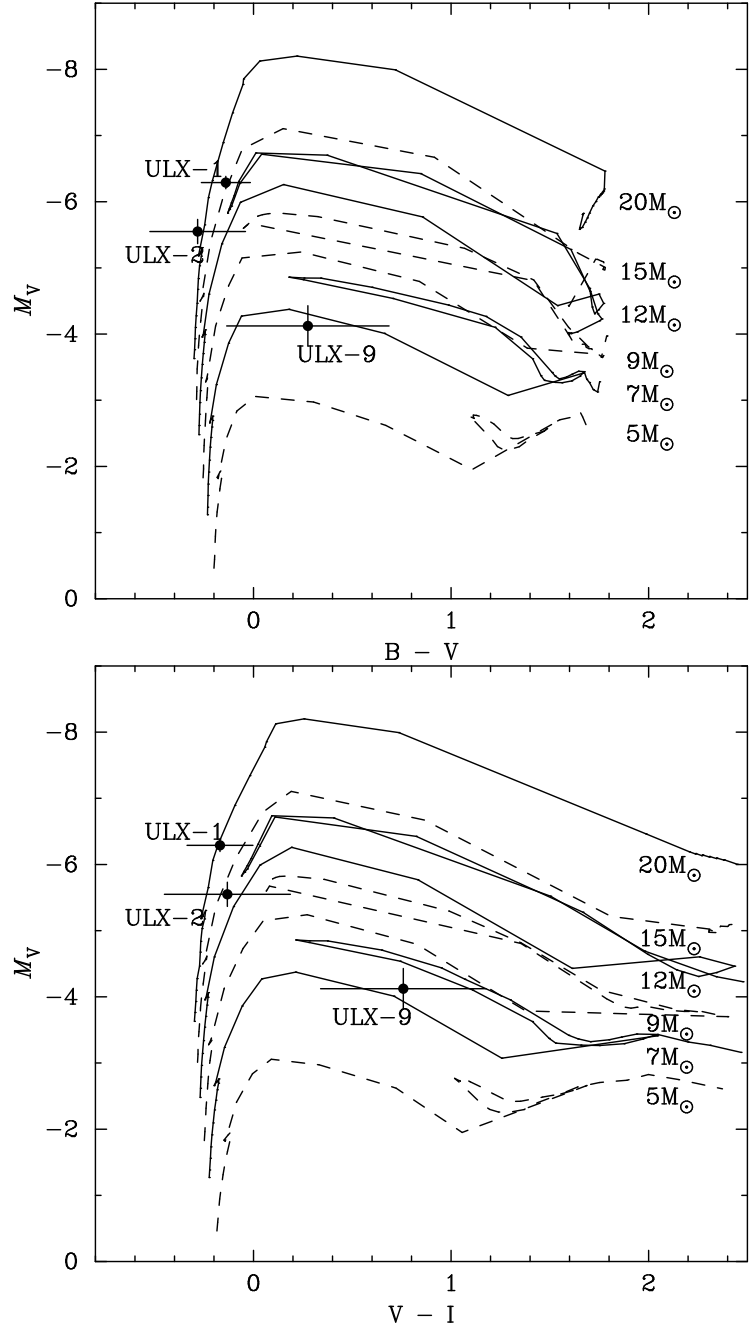

\begin{center}
%\plottwo{fig/fig3a.ps}{fig/fig3b.ps}
\includegraphics[scale=0.5,angle=-90]{f4a.ps}
%\vspace*{10mm}
\includegraphics[scale=0.5,angle=-90]{f4b.ps}
\caption{Locations of candidate counterparts of ULX-1, ULX-2, and
ULX-9 in color-magnitude diagrams.  The stellar evolutionary tracks of
20, 15, 12, 9, and 7$M_{\odot}$ (from top to bottom) of Lejeune \&
Scherer (2001) are shown as solid and dashed lines.
}
\end{center}
\end{figure}

\end{document}